# $R_0$ and the elimination of HIV in Africa: Will *90-90-90* be sufficient?

Brian G. Williams[1] and Eleanor Gouws[2]

1 South African Centre for Epidemiological Modelling and Analysis (SACEMA), Stellenbosch, South Africa
2 UNAIDS Regional Support Team for Eastern and Southern Africa, Johannesburg, South Africa

Correspondence to BrianGerardWilliams@gmail.com

**Abstract**

The Joint United Nations Programme on HIV and AIDS (UNAIDS) has set a new *90-90-90* global target for the coverage of anti-retroviral therapy (ART) to be reached by 2020. This would mean that 90% of all people infected with HIV know their status, 90% of them are on ART and 90% of them will have full viral load suppression. Here we first estimate the case reproduction number, $R_0$, for countries in sub-Saharan Africa and for India using data on the rate at which the prevalence of HIV increased at the start of the epidemic and the life expectancy of people living with HIV who are not on ART. $R_0$ determines the magnitude of the control problem, that is to say, the extent to which transmission must be reduced to eliminate HIV. We show that in sub-Saharan Africa the median value of $R_0$ is 4.6 and in all but five countries $R_0$ is less than 6.3. If the *90-90-90* target is reached, 73% of all those living with HIV will have full viral load suppression. If this is maintained it should guarantee elimination in 70% of all countries in sub-Saharan Africa and will reduce $R_0$ to less than 2 in the remaining 12 countries, making elimination easy to achieve by increasing the availability of other high impact methods of prevention.

## Introduction

The International AIDS Society recommends immediate treatment with anti-retroviral therapy (ART) for all HIV positive people irrespective of their CD4$^+$ cell count.[1] With good compliance this will reduce their viral load to less than 100 cells/μL within about 4 weeks[2] and will reduce their infectiousness to others by 96% or more.[3-5] With high levels of ART coverage it should be possible to reduce the case reproduction number, $R_0$, to less than 1 and eliminate HIV.[6]

Here we use data on the rate of increase of the prevalence of HIV at the start of the epidemics in sub-Saharan African countries to estimate $R_0$ and determine the extent to which transmission must be reduced if HIV is to be eliminated. This enables us to assess the likely impact on the epidemic of reaching the UNAIDS recommended *90-90-90* treatment target.

## Methods

At the start of the epidemic of HIV, in the simplest dynamical model, each infected people infect other people at a *per capita* rate $\beta$ and die at a *per capita* rate $\mu$. The prevalence of infection then increases at a rate $\rho = \beta - \mu$ and the case reproduction number, $R_0$, is given by

$$R_0 = \beta \int_0^\infty e^{-\mu t}\, dt = \frac{\beta}{\mu}. \qquad 1$$

so that

$$R_0 = \frac{\rho + \mu}{\mu} \qquad 2$$

In practice, mortality in people infected with HIV is not constant but increases approximately linearly with time since infection. We wish to find an expression relating $R_0$ to the initial growth rate and the mortality rate for HIV in this more general case.

The best data for survival as a function of the age at infection and the time since infection with HIV are from the CASCADE study[7] (Figure 1). The data in Figure 1 can be fitted to within the statistical errors by a Γ-distribution, Γ(*s*,*k*), where *s* determines the shape and *k* determines the scale of the distribution. Imposing the same shape parameter for all ages gives the fits shown in Figure 1, where the best fit value of the shape parameter is 3.74 (3.66–3.82). The mean value of the survival for Γ(*s*,*k*) is *m* = *sk* and these values are plotted in
Figure 2. The best fit straight line to the data in
Figure 2 has an intercept at age 0 years of 18.67 years and a slope of −0.201.*

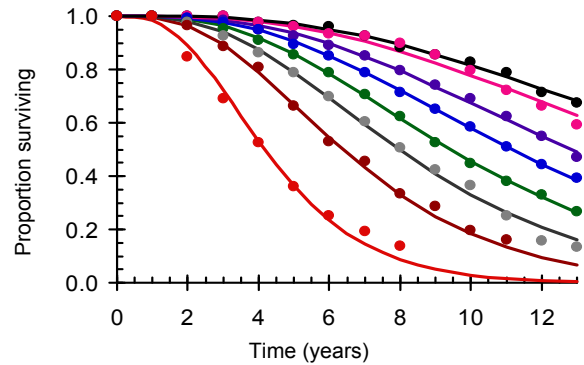

Figure 1. Survival of HIV-positive people as a function of time since infection for people infected at different ages. Orange: 0–5 yrs; brown: 5–15 yrs; grey: 15–25; green: 25–35 yrs; blue: 35–45 yrs; purple: 45–55 years; pink: 55–65 yrs; black: 65–75 yrs.[7]

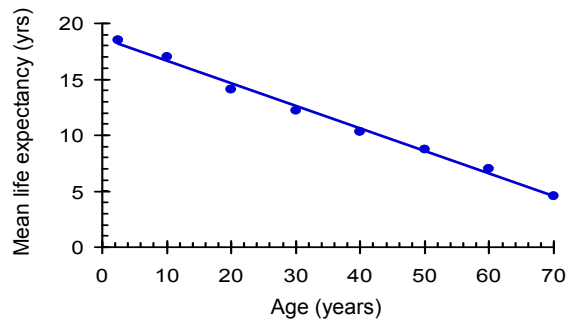

Figure 2. Mean survival for HIV-positive people, not on ART, as a function of age at infection.[7]

---

\* Note that those in the youngest age group (0–5 yrs) were infected through blood-transfusions, not vertically from their mothers at birth.

We first determine the relationship between the epidemic growth rate, $\rho$, the mean survival, $m$, and the rate at which new cases are generated, $\beta$, for different values of the shape parameter $s$. In order to do this we note that for a series of Poisson processes, with $s$ successive states, in which the transition rate from one state to the next is $1/k$, the distribution of time to leaving the last state is $\Gamma(s,k)$. We set up the transition matrix for this process and the principle eigenvector gives the initial rate of increase.

We consider what happens as we increase the shape parameter. In general we have

$$R_0 = \beta \int_0^\infty \Gamma(s,k) t dt \qquad 3$$

$$R_0 = \beta s k \qquad 4$$

so that changing $s$ does not change $R_0$ provided we keep the mean survival time, $sk$, constant. However, if we change $s$ then the relationship between the growth rate, the force of infection and the mean survival, $\beta = \rho + \mu$, no longer holds. For a given value of $R_0$ the growth rate will increase as $s$ increases because there is an increasing time delay between new infections and deaths. If we ignore this correction we will overestimate $R_0$ for a given growth rate. We therefore seek an expression for $R_0$ of the form

$$R_0 = \frac{c\rho + \mu}{\mu} \qquad 5$$

and we choose the correction factor $c$ that gives the best estimates of $R_0$ for different values of $\rho$ and $\mu$.

To estimate $c$ we proceeded as follows. We set up the growth rate matrix for an SI model with either 3 or 4 infectious stages since we want to estimate $c$ for $s = 3.74$. We then set $\mu = 0.1$/year, corresponding to a life-expectancy of ten years, and vary $R_0$ from 1 to 10 in steps of 1 so that $\beta$ varies from 0.1 to 1 in steps of 1 and for each value of $R_0$ and for $s = 3$ and 4 we calculate the principle eigenvalue of the growth rate matrix for a given value of the correction factor $c$. We then regress $R_0$, calculated from Equation 5, against the true value, calculated from Equation 1, by minimizing the maximum absolute value of the error expressed as a percentage of the true value. The mini-max estimate of $c = 0.831$ and the maximum fractional error in the estimates of $R_0$ are all less then 8%. We therefore have as our best estimate of $R_0$

$$R_0 = 1 + \frac{0.831\rho}{\mu} = 1 + \frac{0.576}{d\mu} = 1 + 0.609\frac{n}{d} \qquad 6$$

where we have given alternative expressions for $R_0$ noting that $c = 0.831$, the doubling time $d = \ln(2)/\rho = 0.693/\rho$ and the median survival time, $n$, for the $\Gamma$-distribution is[8]

$$n \approx m\frac{3s-0.8}{3s-0.2} \qquad 7$$

so that for $s = 3.74$ $n \approx 0.946/\mu$.

Finally, we note that the best fit value of $c$ also depends on the value of $\mu$ which varies significantly with the age at infection (Figure 2). We therefore repeat the calculations but with a mean survival of 5 years and 20 years. Equation 6 still gives estimates of $R_0$ that are within 8% of the correct value when $m$ is set to 5 years and to within 12% of the correct value when $m$ is set to 20 years.

## Results

Table 1 gives the estimated doubling times for the prevalence of HIV at the start of the epidemic[9] and Figure 2 shows that the median survival of 30 year old HIV positive people who are not on ART is 12.3 years. Equation 6 then gives the estimated values of $R_0$ (Table 1) and the minimum and maximum values and three percentiles (Table 2).

Table 1. Initial doubling times, $d$, and the corresponding values of $R_0$ for HIV in sub-Saharan African countries and India.

| Country | $d$ (yrs) | $R_0$ | Country | $d$ (yrs) | $R_0$ |
| --- | --- | --- | --- | --- | --- |
| Angola | 1.77 | 4.80 | Lesotho | 1.01 | 7.99 |
| Benin | 1.63 | 5.20 | Liberia | 0.89 | 8.93 |
| Botswana | 1.65 | 5.15 | Madagascar | • | • |
| Burkina Faso | 1.66 | 5.10 | Malawi | 1.95 | 4.62 |
| Burundi | 2.05 | 4.14 | Mali | 1.23 | 6.74 |
| Cameroon | 1.51 | 5.61 | Mozambique | 2.00 | 4.53 |
| CA Republic | 1.69 | 5.02 | Namibia | 1.92 | 4.68 |
| Chad | 3.31 | 2.56 | Niger | 1.89 | 4.74 |
| Congo | 4.31 | 1.97 | Nigeria | 1.17 | 7.03 |
| Côte d'Ivoire | 1.75 | 4.84 | Rwanda | 3.26 | 3.17 |
| Djibouti | 2.06 | 4.13 | Senegal | 3.30 | 3.14 |
| DR Congo | • | • | Sierra Leone | 1.52 | 5.65 |
| Equat. Guinea | 2.73 | 3.10 | Somalia | 5.26 | 2.34 |
| Eritrea | 1.73 | 4.89 | South Africa | 1.47 | 5.80 |
| Ethiopia | 1.55 | 5.46 | Sudan | 5.23 | 2.35 |
| Gabon | 1.93 | 4.40 | Swaziland | 1.96 | 4.60 |
| Gambia | 4.39 | 1.94 | Togo | 1.64 | 5.31 |
| Ghana | 1.81 | 4.68 | Uganda | 3.11 | 3.27 |
| Guinea | 1.94 | 4.38 | UR Tanzania | 2.03 | 4.48 |
| Guinea-Bissau | 2.52 | 3.37 | Zambia | 3.23 | 3.19 |
| India | 1.79 | 4.74 | Zimbabwe | 1.66 | 5.25 |
| Kenya | 1.34 | 6.34 | | | |

Table 2. The minimum and maximum values and the 10%, 50% and 90% percentiles for $R_0$ in sub-Saharan African countries. The right hand column gives the incidence rate ratio, IRR, for those on ART needed to eliminate HIV.

| Percentile | $R_0$ | IRR |
| --- | --- | --- |
| Minimum | 2.34 | 0.43 |
| 10% | 2.55 | 0.39 |
| 50% | 4.57 | 0.22 |
| 90% | 7.31 | 0.14 |
| Maximum | 8.93 | 0.11 |

The median value of $R_0$ is 4.6 and 90% of the estimates are less than 7.3. The five countries with the highest estimated values of $R_0$ are Kenya, Lesotho, Liberia, Mali and Nigeria. In Kenya the early prevalence estimates were based on data from a small but increasing number of antenatal clinics and may not have been representative of the country as a whole. In Lesotho the data are more reliable but the epidemic there was driven in the early stages by migrant labour to the mines in South Africa and may have risen faster than otherwise expected. In Liberia, Mali and Nigeria the peak prevalence was less than 4% so that estimates of early epidemic trends may also be unreliable.

## Discussion

The median value of $R_0$ in sub-Saharan African countries is 4.6 so that to eliminate HIV in half of the countries in sub-Saharan Africa it will be necessary to reduce transmission by 78%; the 90th percentile value of $R_0$ is 7.3 so that to eliminate HIV in 90% of the countries in Africa it will be necessary to reduce transmission by 86%. Reaching the UNAIDS *90-90-90* treatment target for 2020 in which 90% of people infected with HIV know their status, 90% of these are on treatment and 90% of these have full viral load suppression will reduce transmission by 73% and guarantee elimination in 70% of all countries in sub-Saharan Africa and in India. It would also reduce $R_0$ to less than 2 in the remaining 30% of countries.

In several countries in sub-Saharan Africa the prevalence of HIV fell markedly before ART became widely available, suggesting that there must have been a significant reduction in transmission before ART became available.[9] Combined with further increases in the coverage of high impact interventions such as male circumcision and pre-exposure prophylaxis for sex-workers and others at high risk, reaching the UNAIDS target for coverage of ART should almost certainly lead to the eventual elimination of new infections in all of sub-Saharan Africa and in India.

In African countries elimination of HIV will depend on frequent and widespread testing for HIV and high rates of compliance with ART. Other prevention methods including male circumcision, pre-exposure prophylaxis, treating curable sexually transmitted diseases, condom use promotion and behaviour change programmes will not, in themselves, be sufficient to eliminate HIV, but may make an important contribution to early treatment with ART.